\documentclass[twocolumn,prl,twoside,preprintnumbers,floatfix]{revtex4}

\usepackage{amsmath}
\usepackage{amssymb}
\usepackage{graphicx}
\usepackage{dcolumn}
\usepackage{bm}

\def\epsilon{\varepsilon}
\def\theta{\vartheta}
\def\rho{\varrho}

\begin{document}


\title{Comment on ''Copenhagen Interpretation of Quantum Mechanics Is Incorrect''}

\author{Markus Bier}
\email{bier@fluids.mpi-stuttgart.mpg.de}

\affiliation{
   Max-Planck-Institut f\"ur Metallforschung,  
   Heisenbergstra\ss e 3, 
   70569 Stuttgart, 
   Germany \\
}

\affiliation{
   Institut f\"ur Theoretische und Angewandte Physik, 
   Universit\"at Stuttgart, 
   Pfaffenwaldring 57, 
   70569 Stuttgart, 
   Germany
}

\date{September 15, 2005}

\begin{abstract}
It is shown that ''Theorem 1'' of the article ''Copenhagen Interpretation of Quantum
Mechanics Is Incorrect'' by G.-L. Li and V.O.K. Li (see quant-ph/0509089) is false. 
Therefore the assertion expressed in the title of that article is untenable.
\end{abstract}

\maketitle


In Ref. \cite{Li2005}, G.-L. Li and V.O.K. Li attempt to proof ''Theorem 1:
For a probability space $(\Omega,\mathcal{F},P)$, there are values almost everywhere
in $(0,1)$ that the probability measure $P$ cannot take.'' From this assertion, they
deduce non-existence of totally continuous probability measures, i.e., probability
measures with a probability density. Finally, these authors conclude that the 
Copenhagen Interpretation of quantum mechanics, which interprets the squared modulus 
of wave functions $|\psi|^2$ as probability density, would be incorrect. 

But ''Theorem 1'' is obviously false. A trivial counterexample is given by \cite{MTPT}
$\Omega := [0,1]$, $\mathcal{F} := \mathcal{B}\cap[0,1]$, and 
$P := \lambda_{LB}|_{\mathcal{B}\cap[0,1]}$, where $\mathcal{B}$ is the Borel set on 
$\mathbb{R}$ and $\lambda_{LB}$ is the Lebesgue-Borel measure on $\mathcal{B}$:
In this case, it is $P(\mathcal{F}) = [0,1]$.

In the ''Proof'' of ''Theorem 1'', $P(\mathcal{F})$ is inclosed in a set 
$S := \cup_{F\in G(\mathcal{F})}\Phi(F)$, where $\Phi(F)$ is a countable subset
of $[0,1]$ and thus a null-set \cite{Li2005}. Li and Li attempt to show $P(S) = 0$. But,
as $G(\mathcal{F})$ is \emph{un}countable if $\Omega$ is, this statement involves at least 
two flaws: Firstly, $S \in \mathcal{F}$ is \emph{not} evident, i.e. it is \emph{not} guaranteed 
whether $P(S)$ is meaningful either. Secondly, presumably $S \in \mathcal{F}$, from 
$\forall F\in G(\mathcal{F}): P(\Phi(F)) = 0$ one can\emph{not} conclude 
''$P(S) = \sum_{F\in G(\mathcal{F})} P(\Phi(F)) = 0$'' because $P$ is (in general only) 
$\sigma$-additive.

All conclusions drawn from ''Theorem 1'' of Ref. \cite{Li2005} are therefore untenable. 
In particular, the claimed incorrectness of the Copenhagen Interpretation of quantum 
mechanics is unfounded.



\end{document}